\documentclass[reprint,amsmath,amssymb,aps,prb]{revtex4-2}
\usepackage{textcomp}
\usepackage[utf8]{inputenc}
\usepackage{color}
\usepackage{amsmath}
\usepackage{graphicx}
\usepackage{braket}
\usepackage{placeins}
\usepackage{booktabs}
\usepackage{easyReview}

\usepackage[
  bookmarks=false,
  breaklinks=true,
  colorlinks=true,
  citecolor=blue,
  linkcolor=blue,
  urlcolor=blue
]{hyperref}
\usepackage{xr-hyper}

\usepackage{xcolor}

\begin{document}

\title{Interfacial control of magnetism and electron transport in nonisostructural SrRuO$_3$/SrCuO$_2$ heterostructure}

\author{Digbijaya Palai$^{1,2}$}
\thanks{digbijaya.p@iopb.res.in}
\author{B.~Maharana$^{3}$}
\author{P.~Biswal$^{1,2}$}
\author{Shwetha G.~Bhat$^{4}$}
\author{D.~Nayak$^{2,4}$}
\author{D.~Sahoo$^{6}$}

\author{R.~Soni$^{6}$}
\author{K.~Senapati$^{2,4}$}
\author{Z.~Hossain$^{3}$}

\author{D.~Samal$^{1,2}$}
\thanks{dsamal@iopb.res.in}

\affiliation{$^{1}$Institute of Physics, Sachivalaya Marg, Bhubaneswar 751005, India}
\affiliation{$^{2}$Homi Bhabha National Institute, Training School Complex, Anushakti Nagar, Mumbai 400094, India}
\affiliation{$^{3}$Department of Physics, Indian Institute of Technology Kanpur, Kanpur, 208016, India}
\affiliation{$^{4}$ Department of Physics, Indian Institute of Science, Bangalore, 560012, India}
\affiliation{$^{4}$ School of Physical Sciences, National Institute of Science Education and Research (NISER), Bhubaneswar, Jatni, 752050, Odisha, India}
\affiliation{$^{6}$Department of Physical Sciences, Indian Institute of Science Education and Research Berhampur, Berhampur, 760010, India}

\begin{abstract}
Heterointerfaces between structurally dissimilar oxides provide a platform to exploit the interfacial mismatch in the lattice and electronic degrees of freedom to realize emergent functionalities and tunable physical properties with potential applications in oxide electronics. Here, we investigate the magnetic and electronic transport properties of symmetry-mismatched SrRuO$_3$ (SRO)/SrCuO$_2$ (SCO) heterostructure, in which a 4-nm-thick itinerant ferromagnetic metal SRO is interfaced with a planar-type antiferromagnetic insulator SCO, in comparison with a reference SRO (4 nm) film. While both the bare SRO and SRO/SCO bilayer exhibit perpendicular magnetic anisotropy (PMA), the SRO/SCO bilayer shows a pronounced enhancement in the saturation magnetization (M$_S$ $\approx$ 2.7 $\mu_B$/Ru) and effective anisotropy constant (K$_{\mathrm{eff}}$ $\approx$ 2.13 $\times$ 10$^6$ erg/cc) relative to the bare SRO film (M$_S$ $\approx$ 1.3 $\mu_B$/Ru and K$_{\mathrm{eff}}$ $\approx$ 5.25 $\times$ 10$^5$ erg/cc). Interestingly, the low-temperature resistivity upturn below 5 K arising from disorder-induced quantum corrections in bare SRO is strongly suppressed in SRO/SCO,  and it  exhibits Fermi-liquid-like transport ($\rho \propto T^2$) down to 2 K. Analysis of the anomalous Hall effect (AHE) reveals dominant intrinsic Berry-curvature driven electron transport in both samples, with enhanced intrinsic and skew-scattering contributions in the SRO/SCO bilayer, indicating the critical role of the interface in modifying the electronic band structure near the Fermi level. Our results demonstrate that the interface acts as an effective control knob, concurrently enhancing the magnetization, PMA, and intrinsic scattering contribution to the AHE in the SRO/SCO heterostructure, while suppressing the disorder-driven quantum corrections to transport behavior observed in the bare SRO film.

\end{abstract}


\maketitle


\section{Introduction}

Heterostructures designed from transition metal oxide (TMO) have emerged as a versatile material platform to realize emergent phenomena and new functionality absent in their individual constituents. These phenomena arise from interfacial symmetry breaking and the resulting re-organisation of charge, spin, lattice, and orbital degrees of freedom \cite{Zubko2011InterfacePhysics,Ramesh2019EmergentOxideSuperlattices,Trier2022OxideSpinOrbitronics}. Among the broad class of oxide heterostructures, SrRuO$_3$ (SRO) based heterostructures have attracted enormous attention over the years. SRO is an itinerant 4$d$ ferromagnet (T$_c\sim$160 K) with sizable spin-orbit coupling and electron correlations, and its ferromagnetic (FM) metallic ground state can be described by Stoner itinerant model \cite{moriya1985spin,PhysRevB.53.4393,PhysRevB.56.2556}. When integrated with other functional oxides or designed in particular thin film geometry, it unlocks unique functionalities such as strong spin-orbit coupling (SOC) mediated effects including nontrivial electronic topology \cite{Jeong}, canted ferromagnetism and spin dynamics \cite{PhysRevB.105.245107}, and anisotropic in-plane magnetoresistance (MR) \cite{PhysRevB.85.235409}, perpendicular magnetic anisotropy \cite{PhysRevMaterials.5.124403,svc8-dy9j}, topological states \cite{PhysRevB.88.125110,Takiguchi2020WeylSRO,Khetan}, hump-like signal in the Hall effect \cite{Matsuno2016InterfaceDrivenTopologicalHall,Ohuchi2018ElectricFieldHall,Qin2019TopologicalHallSRO}, which are ideal for next-generation spintronics.

The electronic and magnetic properties of SRO is very much sensitive to local structure and electronic environment. Perturbations such as reduced dimensionality, epitaxial strain, and interfacial symmetry breaking can substantially alter the Ru-O-Ru bond network and electronic structure, thereby reshaping the competition among electron correlations, spin-orbit coupling, and magnetic exchange interactions that govern the underlying physical properties. This susceptibility enables interfacial control of electronic and magnetic ground state of SRO through heterostructure design. Indeed, there have been several SRO-based heterostructures exhibit remarkable phenomena such as the control of the anomalous Hall effect (AHE) in SRO/SrIrO$_3$ (SIO) \cite{Ohuchi2018ElectricFieldHall}, increase of T$_C$ in  SRO/SrTiO$_3$ (STO) \cite{PhysRevLett.119.177203}, the emergence of the topological Hall effect (THE) in SRO/SrIrO$_3$ (SIO) \cite{Matsuno2016InterfaceDrivenTopologicalHall}, the interfacial control of ferromagnetism in SRO/BaTiO$_3$ (BTO) \cite{Gu2020InterfacialControlFerromagnetism}, and the modification of magnetic anisotropy and electron transport in SRO/LaFeO$_3$ (LFO) \cite{svc8-dy9j}. However, most of the above studies have focused on interfaces formed between structurally compatible perovskite oxides, where the atomic stacking naturally follows the conventional AO-BO$_2$ sequence along the (001) crystallographic direction.

Despite significant advances in isostructural oxide heterostructures, interfaces between perovskite and non-perovskite oxides remain comparatively less explored. Such nonisostructural interfaces offer distinct opportunities for interfacial engineering, as the structural asymmetry and mismatch in oxygen coordination environment can drive orbital reconstruction and electronic redistribution, giving rise to emergent magnetic and transport phenomena \cite{Zhang2018SymmetryMismatchPMA,Chen2013HighMobility2DEG,PhysRevApplied.12.054016,Jo2020HeterointerfaceBrownmillerite,cspx-cf9j,PhysRevB.109.144423}. Here, we investigate a nonisostrutural oxide interface involving SRO/SrCuO$_2$ (SCO); where SCO is an antiferromagnetic (AFM) insulator. SCO adopts a quasi-two-dimensional lattice (infinite layer) composed of CuO$_2$ planes separated by Sr layers and lacks apical oxygen beyond a critical thickness of 5 unit-cells \cite{PhysRevLett.111.096102,PhysRevB.85.121411,Kuiper2013ControlOxygenSublattice}. The structural dissimilarity at the SRO/SCO interface transforms RuO$_6$ octahedral to a RuO$_5$ square-pyramidal coordination, which favours the Ru $d_{z^2}$ orbital occupancy through the charge transfer from Cu to Ru thereby stabilizing the high spin state of Ru ion ($t_{2g}$: 3$\uparrow$1$\downarrow$, $e_g$: 1$\uparrow$; S = 3/2) \cite{Li2021HighSpinSCO_SRO}. In addition, the planar-type SCO has been used in the heterostructures as an effective means to manipulate the orbital configuration\cite{Liao2019LargeOrbitalPolarization} and enable modulation of the magnetic anisotropy \cite{c6hy-pfxy}. Recently, the theoretical work by Chen et al. on SrCuO$_2$/SrRuO$_3$ heterostructure find that magnetic anisotropy (MA) and Dzyaloshinskii-Moriya interaction (DMI) can be modulated at the at the CuO$_2$-Sr-RuO$_2$ interface resulting from the occupation of the Ru $d_{z^2}$ orbital, inversion symmetry breaking, and orbital hybridization at the interface, paving the way for engineering emergent magnetic and electronic functionalities \cite{PhysRevB.105.214428}. 

Electronic transport measurements reveal interfacial effects through their influence on charge carriers. Herein, we have undertaken a comprehensive experimental study on SRO/SCO interface to examine the effect of the planar-type SCO  layer on the longitudinal and Hall transport. In particular, the magnetic and magnetotransport experiments have been carried out on a 4 nm SRO single layer and an SRO (4 nm)/SCO (24 nm) bilayer grown on SrTiO$_3$ substrates by pulsed laser deposition. From static magnetization measurements, we noticed that the SCO overlayer drives a low-spin ($\approx$1.3 $\mu_B$/Ru) to high-spin ($\approx$2.7 $\mu_B$/Ru) transition of the Ru ion manifested through a pronounced enhancement in the saturation magnetization. This is discussed in the context of possible interfacial orbital reconstruction and charge transfer from Cu to Ru at the SRO/SCO interface in line with the recent report by Li et al. \cite{Li2021HighSpinSCO_SRO} and is supported by the increased carrier concentration in the SRO/SCO bilayer compared to the bare SRO film. The effective magnetic anisotropy along the out-of-plane (OOP) direction is also enhanced in the SRO/SCO bilayer, with (K$_{\mathrm{eff}}$) increasing from ($\approx$ 5.25$\times$10$^5$ erg/cc) in the bare SRO film to ($\approx$ 2.13$\times$10$^6$ erg/cc) in the bilayer. Upon probing the temperature resistivity, the bare SRO film exhibits an upturn at T$\approx$5 K attributed to disorder-induced quantum corrections arising from quantum interference (QI) and/or electron-electron interactions (EEI). Interestingly, the introduction of the SCO layer suppresses this resistivity upturn, and the SRO/SCO bilayer instead displays Fermi-liquid-like transport behaviour down to 2 K. Furthermore, the anomalous Hall effect (AHE) show the nonmonotonic temperature dependence in both bare SRO and SRO/SCO bilayer films. The scaling analysis of the AHE reveals that the intrinsic Berry curvature contribution is amplified in the SRO/SCO bilayer relative to the bare SRO film, possibly due to interfacial modifications of the electronic band structure. Considering both magnetisation and transport measurements on SRO/SCO heterostructure, we show the structurally distinct interface indeed plays a significance role to control the correlated electronic and magnetic properties of SrRuO$_3$ with potential relevance for oxide-based spintronic and topological applications.

\section{EXPERIMENT}
Bare SRO (4 nm) and SRO (4 nm)/SCO (24 nm) bilayer films were deposited on TiO$_2$-terminated SrTiO$_3$ (001) substrates using reflection high energy electron diffraction (RHEED) assisted pulsed laser deposition (KrF excimer laser with $\lambda = 248$ nm) with a repetition rate of 4 Hz (for SRO) and 2 Hz (for SCO). The laser beam was focused to a fluence $\sim$ 1.3 J/cm$^2$ directed onto the SRO and SCO targets. The growth temperature and oxygen pressure were set at 650$^\circ$C and 0.2 mbar, respectively, for both the SRO and SRO/SCO samples. After the completion of the growth, all samples were annealed for 15 minutes, followed by the cooling down to room temperature at a rate of 10$^\circ$C/minute under the same oxygen pressure (0.2 mbar) to eliminate any oxygen deficiency. Further, the SRO/SCO bilayer was capped with STO ($\sim$ 1 nm) using a commercial target to avoid the degradation of the SCO layer. A Rigaku SmartLab high-resolution four-circle X-ray diffractometer was employed to verify the crystallinity and epitaxial quality of the films. The thickness of the samples was estimated through X-ray reflectivity measurements. Magnetic measurements of all samples were performed using a superconducting quantum interference device (Quantum Design SQUID-VSM) magnetometer in the temperature range of 2 to 380 K. Hall-bar devices were fabricated from the thin films using standard photolithography followed by Ar-ion milling. The patterned Hall bars had a channel width of 30 $\mu$m, a longitudinal voltage-probe separation of 740 $\mu$m, and contact pads with dimensions of 200 $\times$ 200 $\mu$m$^2$. Electrical transport measurements were carried out using a closed-cycle cryostat (Cryogenic Ltd.) equipped with a superconducting magnet. All electrical transport measurements were performed using a Cryogen Free Measurement (CFM) system from Cryogenic Ltd., equipped with a cryogen-free superconducting magnet and an integrated variable temperature insert (VTI). The Hall bar devices were mounted on an LCC-20 chip carrier, and electrical contacts were established using a wire bonder (TPT, Model HB-05). The CFM system was integrated with a Lake Shore Cryotronics M81 Synchronous Source Measure (SSM) system consisting of two source modules and two measurement modules. Each measurement module can be configured for DC, AC, or lock-in measurements. A DC current ($I$) of ($\approx$10 $\mu$A) was applied along the $x$-axis of the Hall bar. The actual magnetic field was monitored using a Hall sensor mounted on the VTI radiation shield at the field centre position.

\section{results and discussion}
\subsection{Structural characterization}

\begin{figure*}
\includegraphics[width=\linewidth]{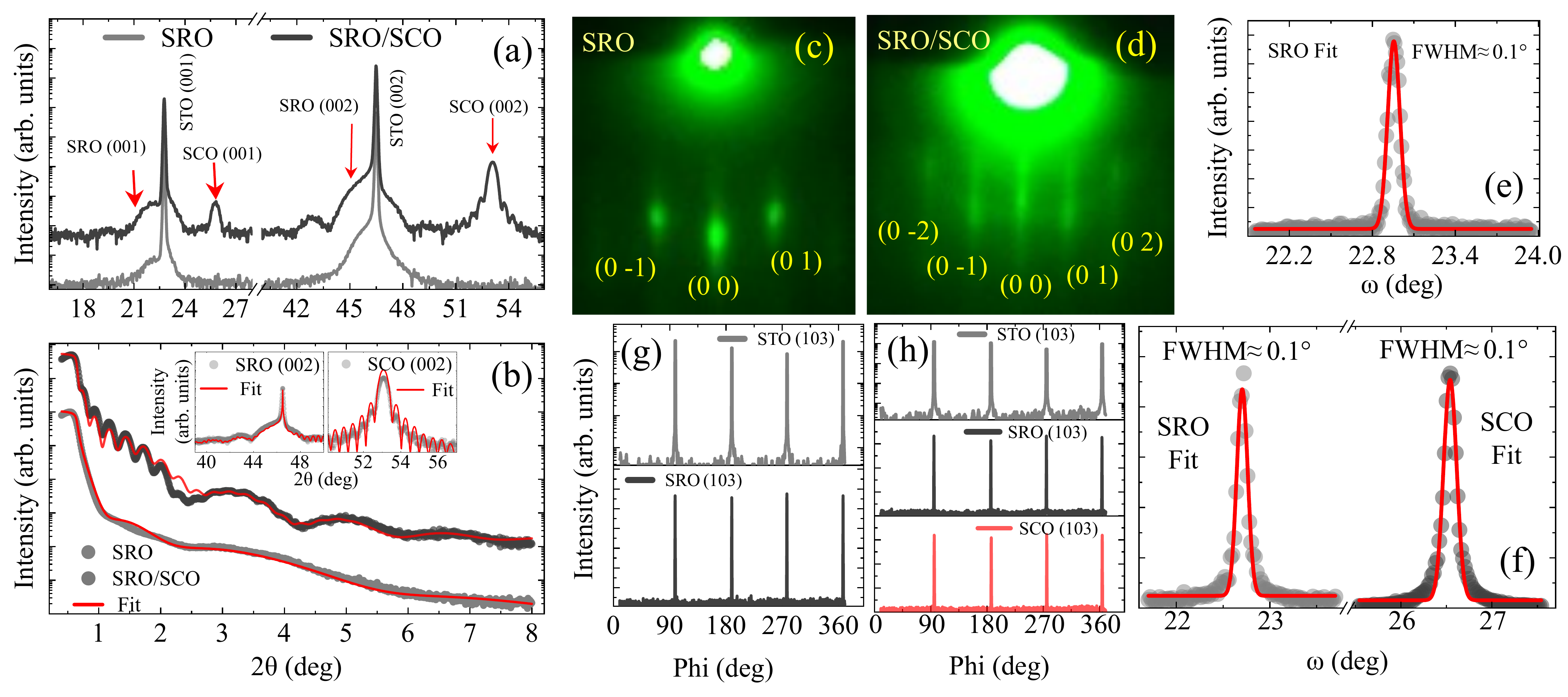}
\caption{(a) X-ray diffraction (XRD) $\theta$–2$\theta$ scans, full range for the SRO single-layer and the SRO/SCO bilayer films. (b) X-ray reflectivity (XRR) data of bare SRO and SRO/SCO bilayer films. (c, d) depicts the RHEED images of bare SRO film, and SRO/SCO bilayer. (e, f) $\omega$-scans (rocking curves) of the bare 4 nm SRO film and  the individual 4 nm SRO and 24 nm SCO layers in the SRO/SCO  bilayer sample. $\phi$-scans of 4 nm SRO layer and STO in bare SRO film (g) and 4 nm SRO, 24 nm SCO, and STO in the SRO/SCO heterostructure (h) showing fourfold symmetry.}
\label{fig:1}
\end{figure*}

Figure ~\ref{fig:1}(a) shows the X-ray $\theta$-2$\theta$ scan of bare SRO and SRO/SCO bilayer films. The distinct (001) and (002) reflections imply the oriented growth of the films and the Laue fringes shown in Fig.~\ref{fig:1}(a) indicate a sharp interface. The out-of-plane lattice constant calculated as 3.947 \AA\ for the SRO layer, which is larger than the bulk value 3.925 \AA, arising from the in-plane compressive strain in the deposited film. The SCO (001) and (002) reflections, respectively, near 2$\theta$ $\approx$ 26.5$^\circ$ and 53$^\circ$ (see Fig.~\ref{fig:1}(a))(thus c-axis lattice parameter$\approx$3.45 \AA), confirming the infinite-layer structure \cite{PhysRevLett.111.096102}. 
The layer-by-layer nature of the growth process was provided by the in situ RHEED pattern [see Figs.~\ref{fig:1}(c) and (d)] at the end of the growth, which showed typical two-dimensional (2D) features for both SRO and SCO layers. The XRR measurements estimate the thickness of bare SRO (4 nm) and SRO (4nm)/SCO (24 nm) by fitting the experimental data as shown in Fig.~\ref{fig:1}(b). The $\theta$-2$\theta$ XRD simulations [Red curves in the inset of Fig.~\ref{fig:1}(b) for SRO, SCO layers] around the (002) reflection match perfectly with the experimental diffraction pattern [grey curves in the same inset], and the thicknesses calculated for SRO and SCO agree with the XRR fit. Rocking curve ($\omega$-scan) measurements show a narrow full width at half maximum (FWHM) of $\sim$0.1$^\circ$, observed for the bare SRO and SRO/SCO bilayer films [Figs.~\ref{fig:1}(e) and (f)], confirming its excellent crystallinity with high structural coherence and minimal mosaicity across the heterostructure. Additionally, $\phi$-scans around the asymmetric (103) reflection were carried out for the bare SRO and SRO/SCO bilayer. As shown in Fig.~\ref{fig:1}(g) and (h), all layers display a clear fourfold symmetry indicating cube on cube epitaxial growth of films on the substrate. For clarity, the SrRuO$_3$ (4nm) single layer and the SRO (4 nm)/SrCuO$_2$ (24 nm) bilayer are hereafter referred to as SRO and SRO/SCO, respectively.


\subsection{Magnetism and MR}

\begin{figure}[!ht]
\centering
\includegraphics[width=1\linewidth]{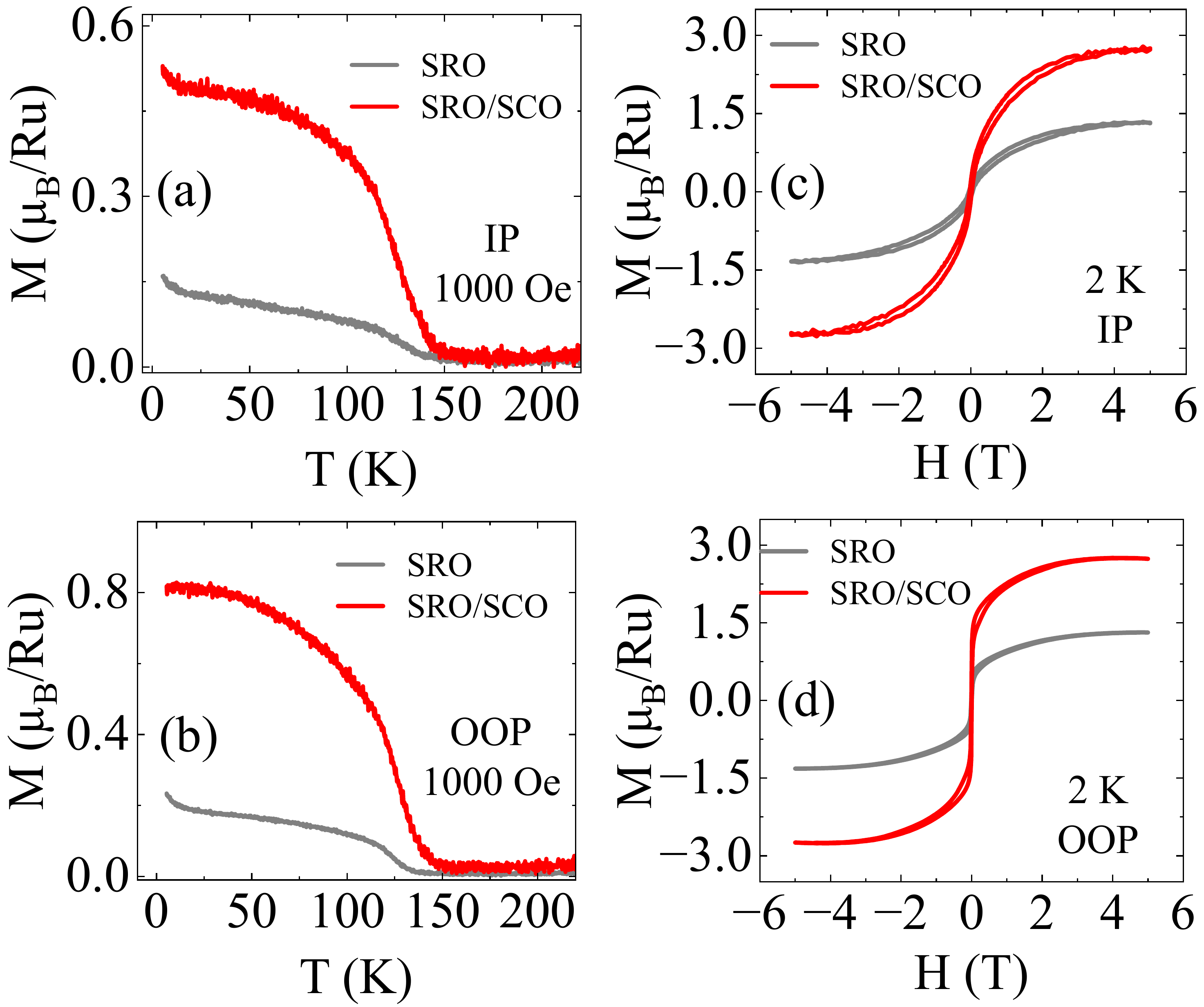}
\caption{M(T) plots for the SRO single layer recorded along the IP (a) and OOP (b) configurations. M(H) loops measured along the IP (c) and OOP (d) orientations for the bare SRO and the SRO/SCO bilyer films at 5 K}
\label{fig:2}
\end{figure}

Figs. ~\ref{fig:2}(a) and ~\ref{fig:2}(b) show the temperature-dependent in-plane (IP) and out-of-plane (OOP) magnetization M(T) measured under an applied field of 1000 Oe for bare SRO and SRO/SCO bilayer films. The Curie temperatures (T$_C$) of the bare SRO and SRO/SCO samples are 135 K and 140 K, respectively. The suppression of T$_C$ relative to bulk SRO (T$_C$ $\approx$ 160 K) in both samples could be attributed to finite-size or strain effects \cite{PhysRevB.92.064402,PhysRevB.88.144412}. Notably, introducing the SCO overlayer results in a measurable enhancement of T$_C$, pointing to a significant interfacial effect. This is further corroborated by the magnetic hysteresis M(H) loops shown in Figs.~\ref{fig:2}(c) and (d), which reveal a substantial increase in saturation magnetization: M$_S$ $\approx$ 1.3 $\mu_B$/Ru for bare SRO and M$_S$ $\approx$ 2.7 $\mu_B$/Ru for the SRO/SCO bilayer. Since magnetism in SrRuO$_3$ is itinerant and governed by the Ru (4$d^4$) electronic configuration, the enhanced magnetisation in the SRO/SCO bilayer can be understood in terms of an interfacial charge transfer that modifies the Ru spin state. The infinite-layer structure of SCO, consisting of alternating Sr$^{2+}$ and CuO$_2^-$ planes, can create a polar discontinuity when interfaced with the RuO$_2$ layer of SRO \cite{Linnanao}. The resulting intrinsic built-in electric field could drive electronic reconstruction and charge transfer from the CuO$_2$ layer toward the adjacent RuO$_2$ layer \cite{Linnanao}. In support of this picture, we estimated the carrier concentration ($n_e$) is found to be higher in the SRO/SCO bilayer than in the bare SRO (see  Fig.S1 in supplemental material (SM)), pointing toward the charge transfer from Cu to Ru across the interface. If such charge transfer occurs, the additional electronic charge on Ru could promote a transition from the low-spin configuration ($t_{2g}$: 3$\uparrow$1$\downarrow$, $e_g$: 0; S = 1), corresponding to a theoretical maximum moment of 2 $\mu_B$/Ru, toward a high-spin state ($t_{2g}$: 3$\uparrow$1$\downarrow$, $e_g$: 1$\uparrow$; S = 3/2) with an enhanced theoretical moment of 3 $\mu_B$/Ru, consistent with the enhanced magnetization observed in the SRO/SCO bilayer. Such interface-driven Ru spin-state transitions in SRO/SCO heterostructures have also been explicitly reported by Li et al. \cite{Li2021HighSpinSCO_SRO}, further in line with this interpretation. The enhancement in T$_C$ can be understood within the Stoner model, given that SRO is an itinerant ferromagnet satisfying the criterion IN(E$_F$) $>$ 1 \cite{moriya1985spin}, where I is the effective electron-electron interaction and N(E$_F$) is the non-magnetic density of states at the Fermi level E$_F$. According to this model, T$_C$ $\propto$ [1 $-$ 1/IN(E$_F$)]$^{1/2}$, so any increment in N(E$_F$) could raise T$_C$ and since the enhanced T$_C$ is observed in the SRO/SCO bilayer, it is reasonable to argue that the SRO/SCO interface appears to increase the N(E$_F$) near the Fermi level in the bilayer.

Further, the specific interfacial stacking geometry plays a decisive role in enabling the high-spin state. In the SRO/SCO bilayer, SCO layer is grown on the top of SRO deposited on a TiO$_2$ terminated STO substarte, the interface is expected to follow a RuO$_2$-Sr-CuO$_2$ stacking sequence along the c-axis, in contrast to the RuO$_2$-SrO-RuO$_2$ periodicity maintained throughout bare SRO grown on STO. This structural modification promotes an elongation of the c-axis through the formation of RuO$_5$ square-pyramidal coordination environment at the interface. The resulting crystal field effect is expected to lower and stabilize the Ru 4d$_{z^2}$ orbital, facilitating electron acceptance from Cu and stabilizing the high-spin state. In the bare SRO film, the RuO$_6$ octahedral coordination preserves the low-spin state of the Ru ion.

\begin{figure}[!ht]
\centering
\includegraphics[width=1\linewidth]{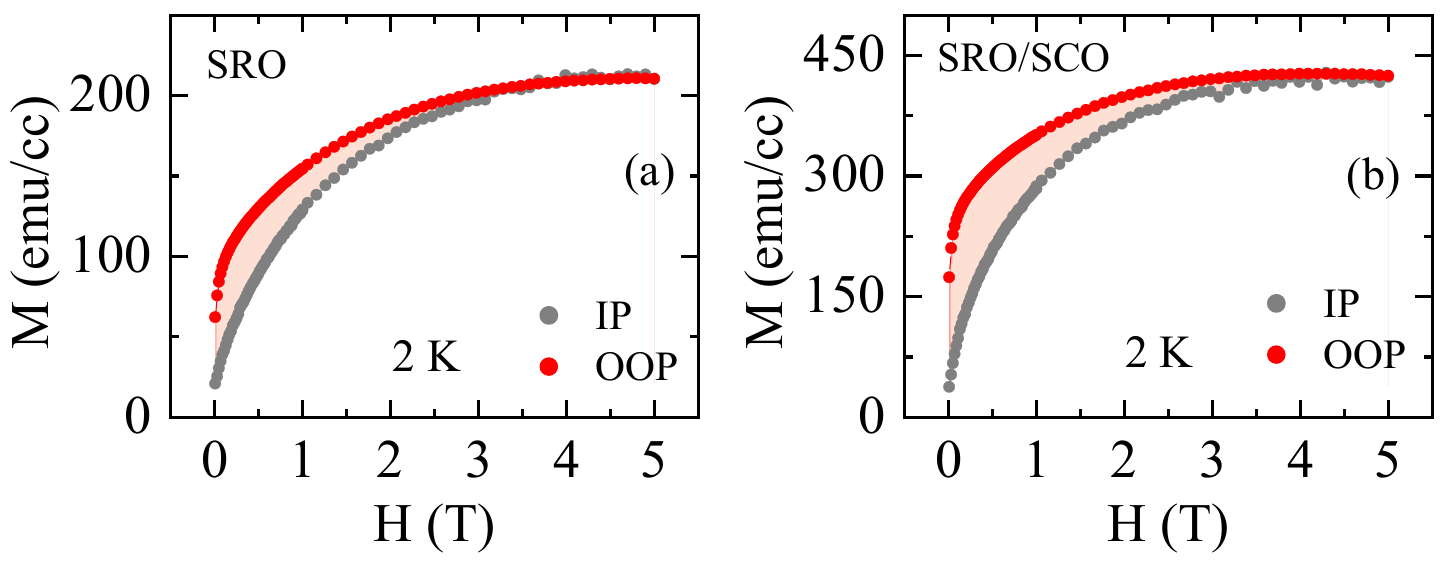}
\caption{Area under the loop in IP and OOP curve in the first quadrant from zero to saturation field (H$_S$) for (a) the bare SRO film and (b) SRO/SCO bilayer at 5 K.}
\label{fig:3}
\end{figure}

Moreover, both the bare SRO and SRO/SCO bilayer exhibit perpendicular magnetic anisotropy (PMA), as evident from the M(T) and M(H) loops in Fig.~\ref{fig:2}. To qualitatively assess how the SCO overlayer influences the magnetic anisotropy of SRO, we examine the area enclosed between the IP and OOP M(H) curves in the first quadrant, from zero field up to the respective saturation fields at 2 K. For the bare SRO film, saturation is reached at H$_S$ $\approx$ 4.1 T (IP) and $\approx$ 3.6 T (OOP), while in the SRO/SCO bilayer these values shift to $\approx$ 3.9 T and $\approx$ 2.9 T, respectively [see ~\ref{fig:3}(a) and (b)]. The simultaneous increase in the IP saturation field and reduction in the OOP saturation field indicates that the OOP orientation is more strongly preferred in the bilayer, signifying an enhanced PMA. To quantify this, we estimate the effective magnetic anisotropy constant K$_{\mathrm{eff}}$ = $\frac{1}{2}$M$_S$H$_a$, where M$_S$ is the saturation magnetization and H$_a$ is the anisotropy field, defined here as the difference in saturation fields between the IP and OOP directions. The resulting K$_{\mathrm{eff}}$ is nearly four times larger in the SRO/SCO bilayer (K$_{\mathrm{eff}}$ $\approx$ 2.13 $\times$ 10$^6$ erg/cc) compared to the bare SRO film (K$_{\mathrm{eff}}$ $\approx$ 5.25 $\times$ 10$^5$ erg/cc), providing clear evidence that the interfacial effect with the addition of SCO layer substantially strengthens the PMA.

\begin{figure}[!ht]
\centering
\includegraphics[width=0.8\linewidth]{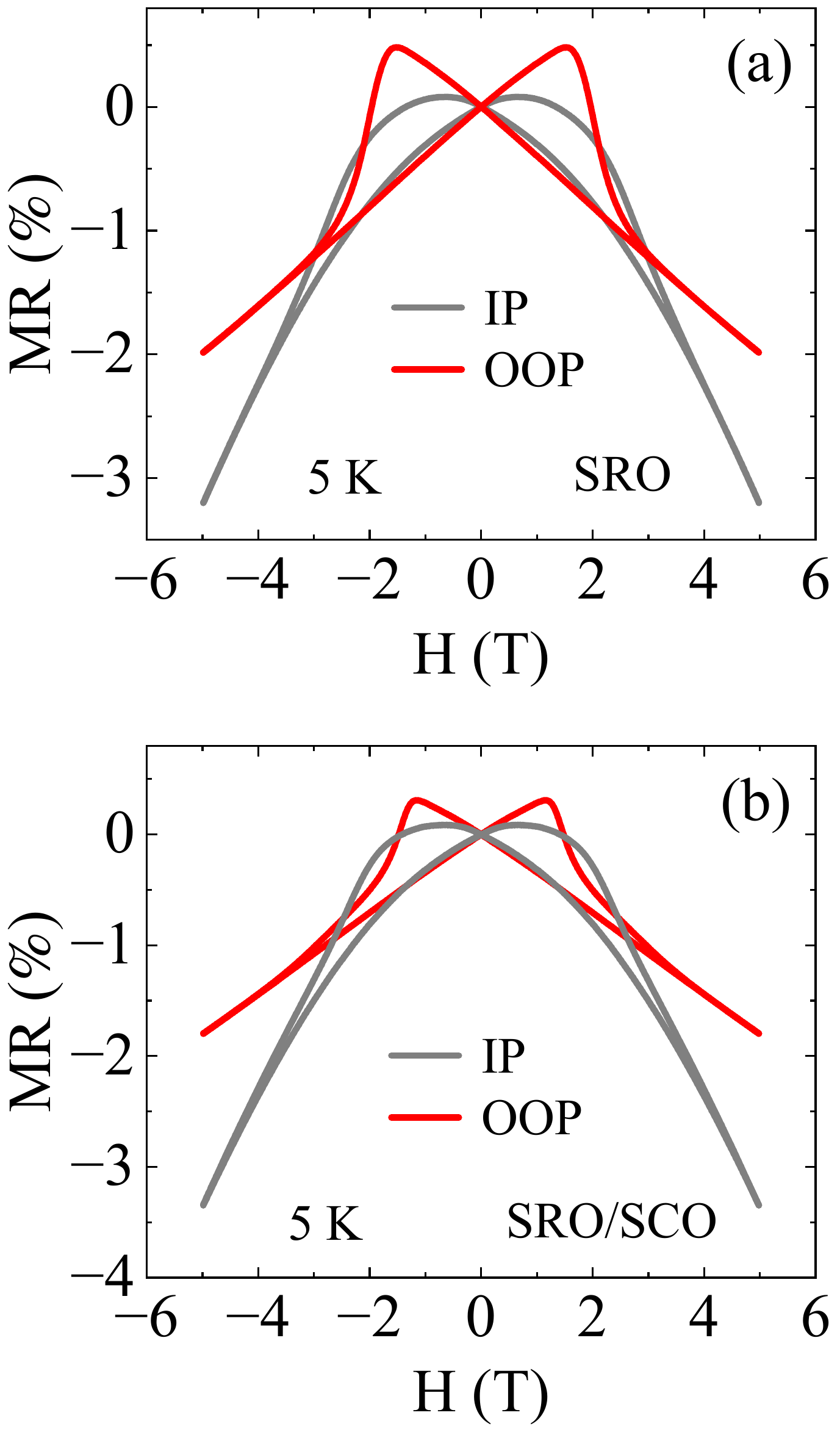}
\caption{Magnetoresistance vs field measured along OOP and IP directions at 5 K for the bare SRO film (a) and the SRO/SCO bilyer (b).}
\label{fig:4}
\end{figure}

As magnetoresistance (MR) is sensitive to spin-dependent scattering and magnetic moment alignment, the field-dependent MR measurements provide an additional probe of magnetic anisotropy. We therefore next discuss the MR behaviour of both the bare SRO and SRO/SCO bilayer films. Figs.~\ref{fig:4}(a) and (b) presents the magnetoresistance MR (\%) = $\{[\rho(H) - \rho(0)]/\rho(0)\} \times 100$ for both the bare SRO film and the SRO/SCO bilayer, measured with the field applied along the OOP and IP directions at 5 K. In both samples, a positive MR at low fields crosses over to a negative MR at higher fields, consistent with earlier reports \cite{svc8-dy9j,Shen2015ThicknessDependentMIT}. Both samples also exhibit butterfly-shaped (along OOP direction, Fig.~\ref{fig:4}(a)) and parabolic (along IP direction, Fig.~\ref{fig:4}(b)) MR hysteresis loops, a characteristic signature of FM ordering. In both films, the relatively sharper low-field switching and the tendency toward earlier saturation in the OOP MR configuration, compared to the more gradual field dependence observed for the IP configuration, suggest that the magnetization responds more readily to the OOP magnetic field. Thus, the MR response is clearly anisotropic, with the OOP direction confirming the easy axis in both bare SRO and the SRO/SCO bilayer, reflecting the dependence of MR on spin-dependent scattering and the orientation of magnetization relative to the applied field. The hysteresis closing field is noticeably smaller in the SRO/SCO (H$_{cl}$ $\approx$ 2.65 T) bilayer than in bare SRO (H$_{cl}$ $\approx$ 2.95 T), indicating that a stronger field is required to achieve complete magnetic alignment in the bare SRO film. This agrees with the enhanced PMA in the SRO/SCO bilayer relative to the bare SRO film from the magnetization measurements.

\subsection{Electron transport}
\begin{figure}[!ht]
\centering
\includegraphics[width=0.8\linewidth]{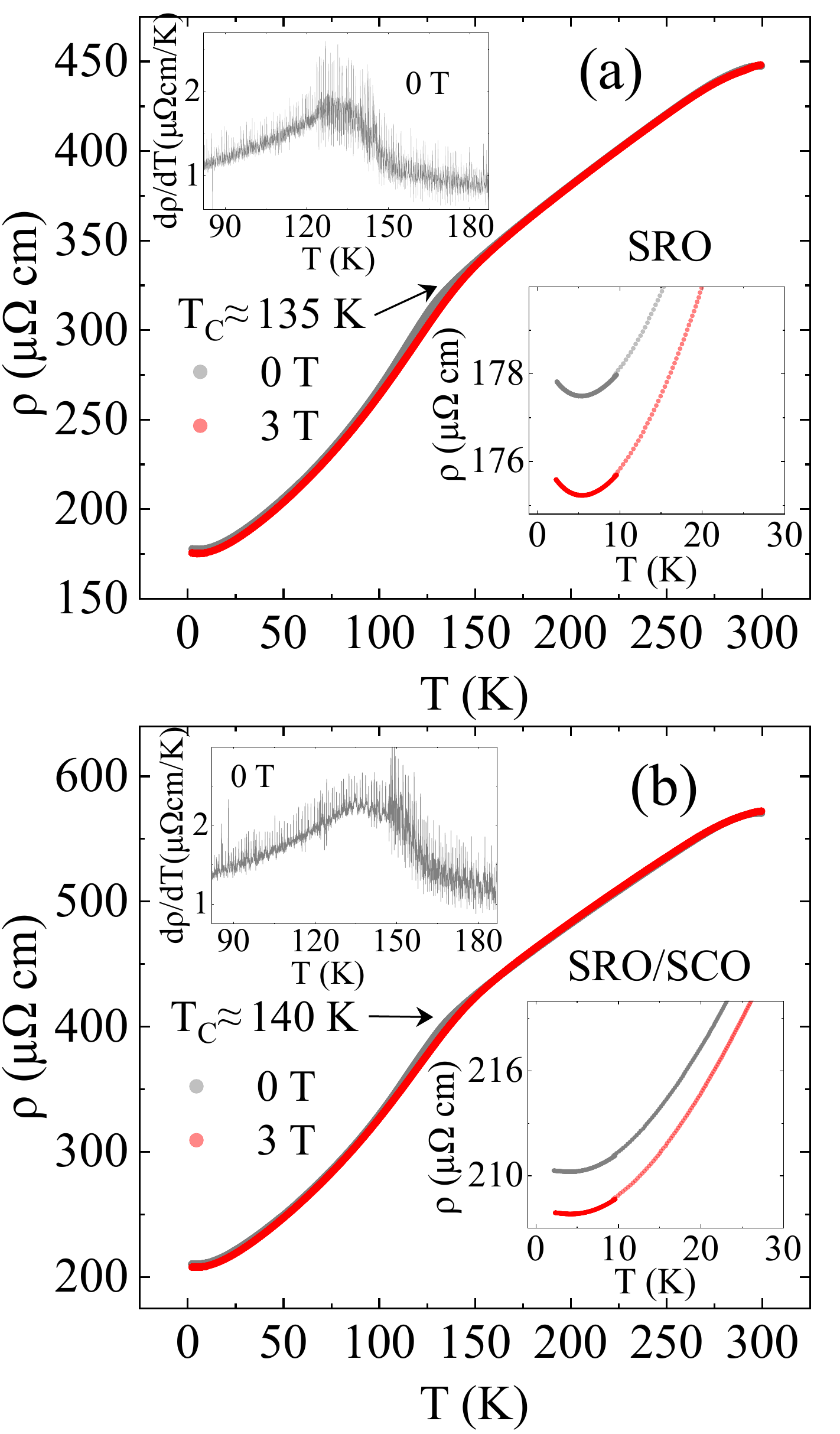}
\caption{Temperature dependence of the resistivity of the bare SRO film (a) and SRO/SCO bilayer (b) in the presence of different magnetic fields (0 T and 3 T). Insets: (upper) T$_C$ determined by the dR/dT of the zero-field $\rho(T)$ curves, (lower) zoom of the low-temperature region of (a) and (b).}
\label{fig:5}
\end{figure}
We now examine the effect of the SCO overlayer on the electronic transport properties of SRO. Figs.~\ref{fig:5}(a) and ~\ref{fig:5}(b), respectively, show the temperature-dependent resistivity [$\rho(T)$] curves for the bare SRO and SRO/SCO bilayer measured under the fields of 0 T and 3 T. The kink observed in the zero-field resistivity curves around 135 K and 140 K consistent with the magnetic ordering temperatures of the bare SRO film and the SRO/SCO bilayer, respectively, obtained from the magnetic measurements. For clarity, the first derivative of resistivity with respect to temperature (d$\rho(T)$/dT), shown in the upper insets of Figs.~\ref{fig:5}(a) and ~\ref{fig:5}(b), show the peaks around the transition temperatures. For the low temperature regime (2 K $<$ T $<$ 25 K), while the bare SRO film exhibits the resistivity upturn below $\approx$ 5 K, apparently this feature is absent in the SRO/SCO bilayer. The possible origins for this behaviour could be attributed to (a) the Kondo effect, (b) quantum interference (QI), and/or enhanced electron-electron interactions (EEI). 

Although Kondo scattering is known to produce a resistivity minimum even in magnetically ordered systems \cite{PhysRevB.99.115135,PhysRevB.73.224410}, its relevance here is questionable. SRO is an itinerant ferromagnet in which the conduction electrons themselves carry the magnetic moment through hybridization between Ru $t_{2g}$ and O 2$p$ orbitals. The formation of isolated localized moments, a prerequisite for the Kondo effect, is therefore unlikely unless the conduction bandwidth is narrowed by A-site substitution \cite{Shepard1997ThermodynamicProperties}, which is not the case here. Moreover, the magnetoresistance (MR) data offer a more definitive test. The Kondo effect typically produces a negative MR, as an applied field suppresses spin-flip scattering from magnetic impurities \cite{PhysRevB.99.115135,PhysRevB.73.224410}, whereas as shown Figs.~\ref{fig:4}(a) and ~\ref{fig:4}(b), the bare SRO film exhibits positive MR at low fields and negative MR at higher fields, a crossover that rules out a purely Kondo-driven upturn \cite{PhysRevB.72.014457,svc8-dy9j}.

\begin{figure}[!ht]
\centering
\includegraphics[width=0.8\linewidth]{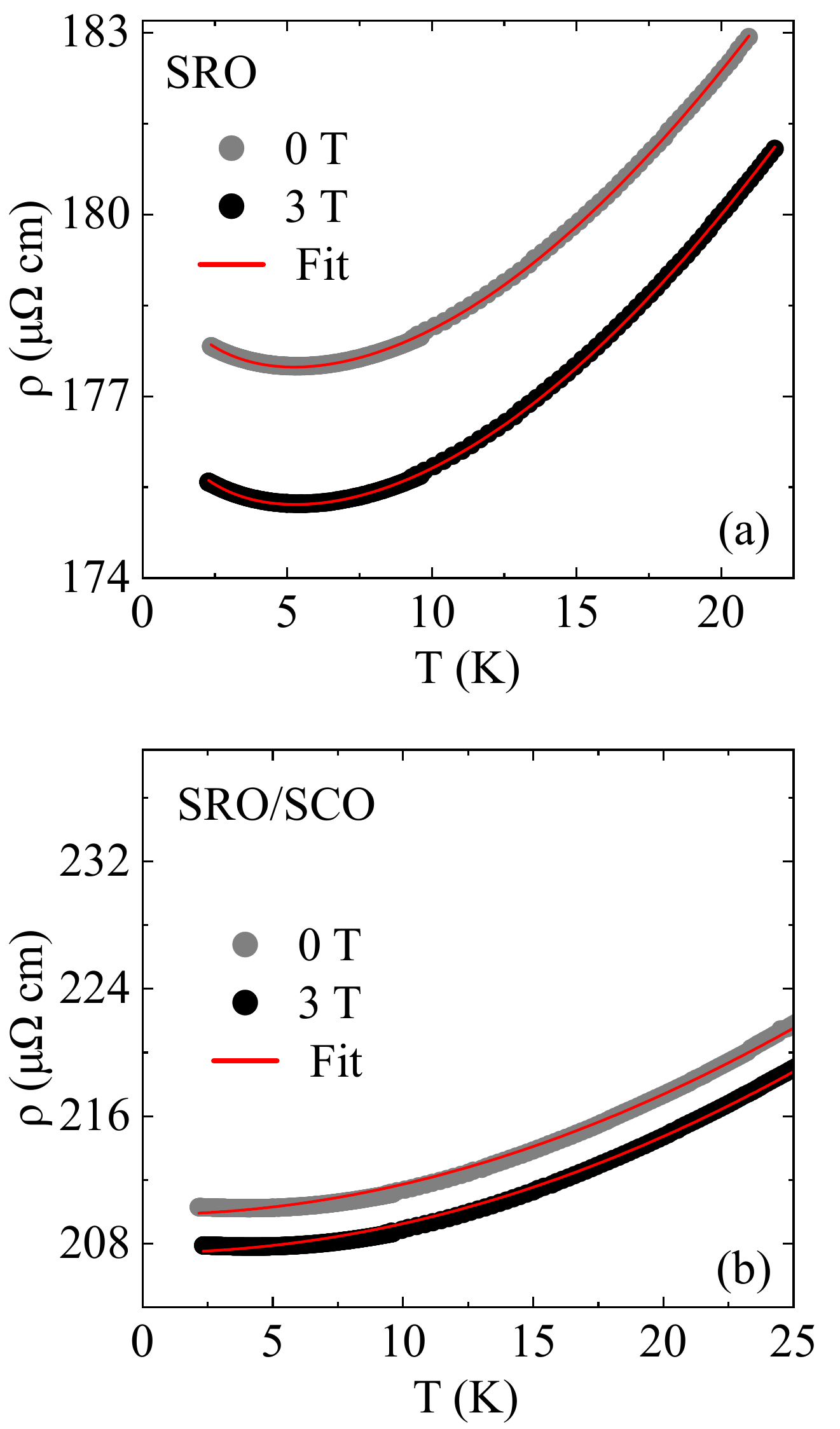}
\caption{Low-temperature (2 K $<$ T $<$ 25 K) resistivity curves fitted with Eq.~\eqref{eq:1} and Eq.~\eqref{eq:2} for the bare SRO film (a) and SRO/SCO bilayer (b) in the presence of different magnetic fields (0 T and 3 T). The scattered symbols correspond to the experimental data, and the red solid lines are the fittings to the Eq.~\eqref{eq:1} and Eq.~\eqref{eq:2}}
\label{fig:6}
\end{figure}

The low temperature resistivity in SRO is fitted [Fig.~\ref{fig:6}(a)] by considering the electron-phonon scattering, electron-electron scattering and QI and /or EEI effects (Eq.~\eqref{eq:1}) [Ref],  
\begin{equation}
\rho(T) = \rho_0 + \rho_{\mathrm{in}}T^5 + \rho_{\mathrm{el}}T^{2} - \rho_l \ln T
\label{eq:1}
\end{equation}
where $\rho_0$ is the residual resistivity, the second term accounts for inelastic scattering from electron-phonon and electron-magnon interactions, the third term represents the electron-electron scattering- a hallmark of Fermi liquid (FL) behaviour and the fourth term corresponds to the QI and/or EEI effect. The coefficient of each term obtained from the fitting is listed in Table 1 in SM. From the above fitting, we understand that the upturn in bare SRO film is associated with the QI and/or EEI effect. 

Strikingly, the low temperature-dependent resistivity upturn is absent in the SRO/SCO bilayer and the temperature dependence of resistivity at low temperatures is well described by the phenomenological model \cite{svc8-dy9j,PhysRevB.99.115135},
\begin{equation}
\rho(T) = \rho'_0 + \rho'_{\mathrm{in}}T^5 + \rho'_{\mathrm{el}}T^2
\label{eq:2}
\end{equation}
where the first term ($\rho'_0$) represents the residual resistivity, second term attributes to the presence of electron-phonon and electron-magnon interactions and the last term corresponds to the delocalised electron-electron scattering mechanism. As shown in Fig.~\ref{fig:6}(b), Eq~\eqref{eq:2} fits the SRO/SCO bilayer $\rho(T)$ data well, with the extracted coefficients listed in Table 2 in SM. It is found that the T$2$term dominates over the inelastic contribution in the given temperature.  Further, it is interesting to note that although SRO/SCO bilayer exhibits a higher resistivity than that of bare SRO film, the upturn is suppressed in the bilayer. This indicates that the upturn is not directly correlated with the magnitude of resistivity. The absence of upturn in $\rho(T)$ of the SCO/SRO bilayer suggests a modification of quantum transport behaviour possibly arising from interface induced modification of electronic structure and/or scattering mechanisms. It has been established that disorder in SRO modifies the density of states (DOS) at the Fermi level (N(E$_F$)) which in turn promotes carrier localization, since the WL correction scales as ~1/(N(E$_F$)) \cite{PhysRevB.63.052403,RevModPhys.57.287,Mott1993ConductionNonCrystalline}. In the magnetism section, we indicated the possibility of enhanced N(E$_F$) in the SRO/SCO bilayer to account for the observed change in magnetisation. Such an increase in N(E$_F$) is expected to suppress the WL contribution in the SRO/SCO bilayer. In essence, the SRO/SCO interface acts as an electronically active platform that reduces the disorder present in the bare SRO layer and modifies its correlated electron transport behaviour. For completeness, the high-temperature transport analysis (25 K $<$ T $<$ 130 K), including the fitting process and corresponding fitting coefficients for both the bare SRO film and the SRO/SCO bilayer, is presented in the section S3 of SM. In this temperature regime, the transport behaviour of both samples is found to be the same, with the electron-electron scattering mechanism dominant with the SRO/SCO bilayer exhibiting only a marginal enhancement in the elastic scattering contribution compared to bare SRO.

\subsection{AHE}

\begin{figure}[!ht]
\centering
\includegraphics[width=1\linewidth]{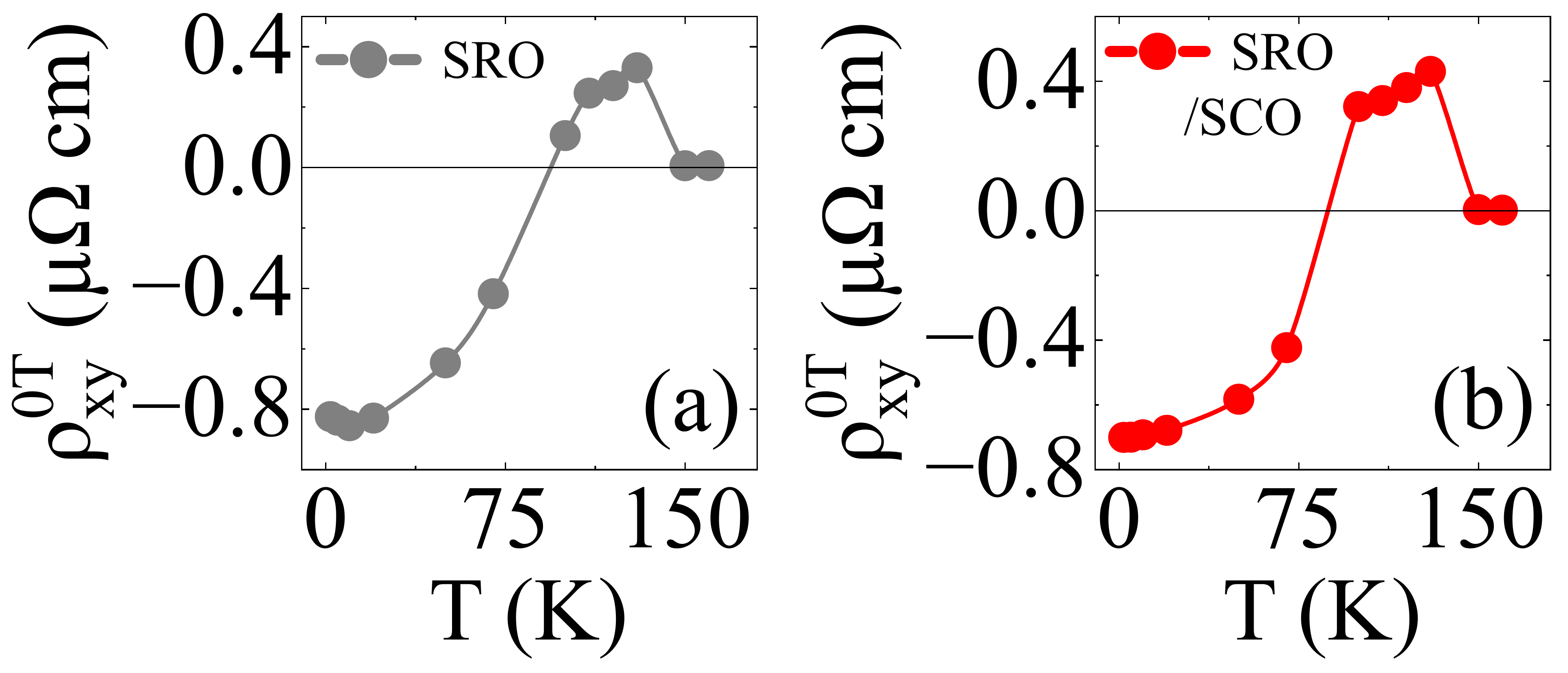}
\caption{Temperature-dependent anomalous Hall resistivity showing the polarity change in the Hall response for the bare SRO film (a) and the SRO/SCO bilayer (b). The solid lines are guides to the eye.}
\label{fig:7}
\end{figure}

\begin{figure}[!ht]
\centering
\includegraphics[width=1\linewidth]{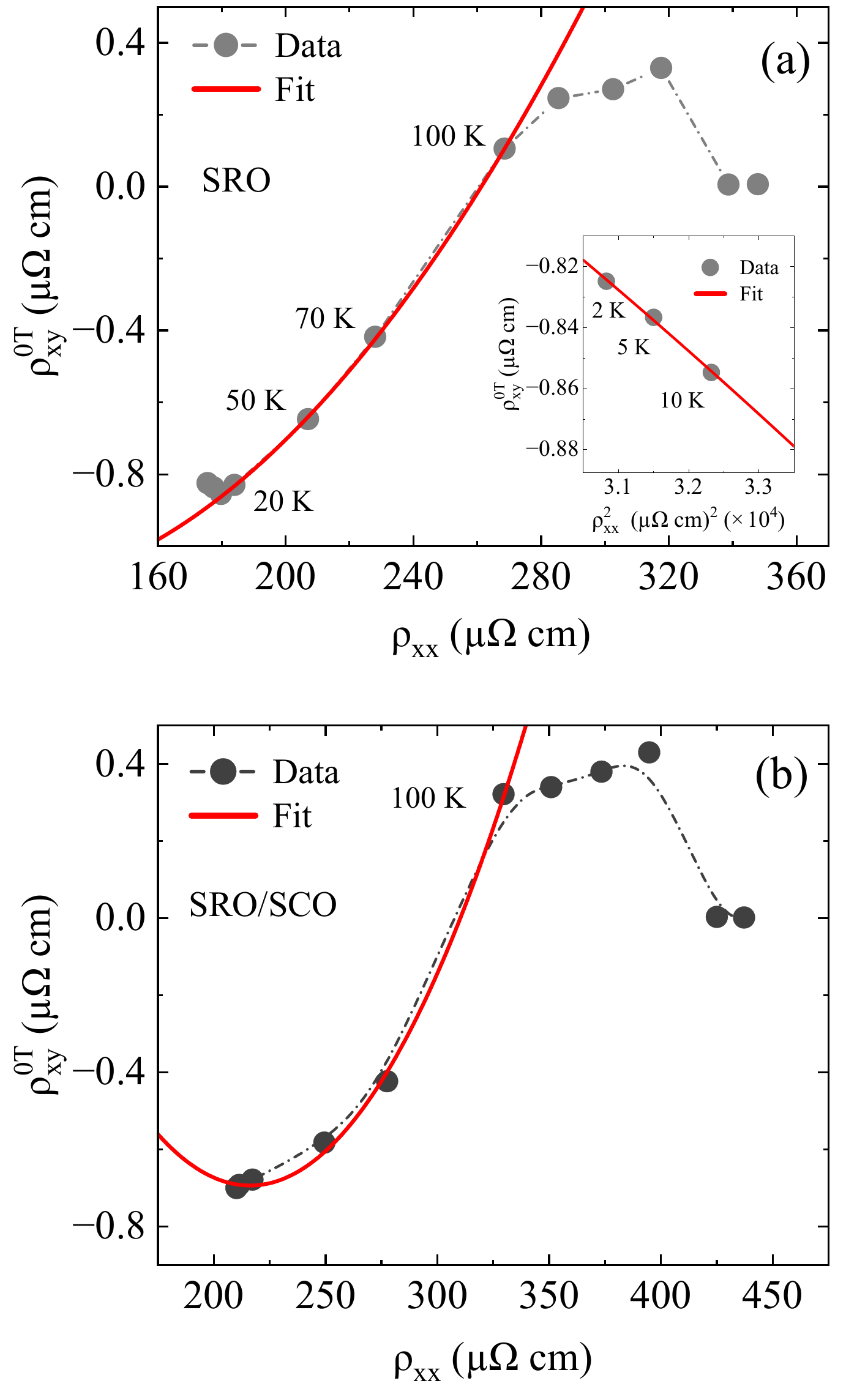}
\caption{Scaling relation between $\rho^{0T}_{xy}$ versus  $\rho_{xx}$ for the bare SRO film (a) and the SRO/SCO bilayer (b), suggesting the presence of both intrinsic and extrinsic skew-scattering contributions, with the intrinsic mechanism dominating. Inset: $\rho^{0T}_{xy}$ fitted only with $\rho_{xx}^{2}$ in the low-temperature regime of the bare SRO film. The dot-dash-dot lines are guides to the eye.}
\label{fig:8}
\end{figure}

To understand the influence of the SCO layer on the anomalous Hall transport of the SRO associated to the change in magnetization, we examine the AHE in the bare SRO film and the SRO/SCO bilayer. Figure S3(a) and (b) in SM, frespectively, show the field-dependent Hall resistivity for the SRO and SRO/SCO across the temperature range 2-130 K, after subtracting the ordinary Hall contribution from the total transverse resistivity. Notably, a sign change in the anomalous Hall resistivity is observed near 100 K for both bare SRO and SRO/SCO bilayer. In a FM, the Hall resistivity is expressed as,
\begin{equation}
\rho_{xy} = \rho_{\mathrm{OHE}} + \rho_{\mathrm{AHE}} = R_0H + R_SM_Z
\label{eq:3}
\end{equation}
where $R_0$ is the ordinary Hall coefficient that depends on the majority charge carrier density, $H$ is the applied magnetic field normal to the film plane, $R_S$ denotes the AHE coefficient and $M_Z$ is magnetization along $H$ direction. The anomalous Hall resistivity $\rho^{0T}_{xy}$ is determined by extrapolating $\rho_{xy}(H)$ to zero field at different temperatures ranging from 2-160 K for both samples, as shown in Fig.~\ref{fig:7}(a)and ~\ref{fig:7}(b). The monotonic temperature dependence of $\rho^{0T}_{xy}$ suggests contributions from both intrinsic and extrinsic mechanisms, consistent with earlier reports \cite{PhysRevB.84.174439,PhysRevMaterials.7.054406}. To examine the contributions of different scattering mechanisms to the AHE, we employ the scaling relation given by Eq. (4) to fit the transverse resistivity data;
\begin{equation}
\rho^{0T}_{xy} = \rho_{xx0} + \alpha\rho_{xx} + \beta\rho_{xx}^{2}
\label{eq:4}
\end{equation}
where $\rho_{xx0}$, $\rho_{xx}$, $\alpha$ and $\beta$ represent the residual resistivity, longitudinal resistivity, skew-scattering, and side-jump/intrinsic contributions, respectively. Because $\rho^{0T}_{xy}$ changes sign in both samples, the quadratic term in Eq. (4) term is reasonably associated with the intrinsic contribution linked to the Berry curvature rather than to side jump. Further, from Fig. S4(a) and (b) in SM, the temperature (T) dependence of $\rho^{0T}_{xy}$ to that of the magnetization M$_S$(T) are presented for the bare SRO and SRO/SCO bilayer, where $\rho^{0T}_{xy}$ after the initial increase, follows a decrease, then increases and changes sign (becoming positive), capturing the non-monotonous behaviour of $\rho^{0T}_{xy}$. These features are basic to the systems where the AHE is attributed to the intrinsic mechanism \cite{Fang}. In addition, the AHC $\sigma_{xx}$ $\approx$ 10$^4$ $\Omega^{-1}$ cm$^{-1}$ places both samples in the moderately dirty regime \cite{Yang2020GiantUnconventionalAHE,PhysRevB.79.014431,Fujishiro2021GiantAnomalousHall}, where intrinsic scattering is expected to govern the AHE \cite{Roy2023OriginTopologicalHall}. On this basis, the analysis proceeds by considering both skew scattering and the intrinsic Berry curvature mechanism.
The scaling relation (Eq.~\eqref{eq:4}) describes the SRO/SCO bilayer data well in the temperature range (2-100 K) as shown in Fig.~\ref{fig:8}(b). For the bare SRO film, however, the low-temperature data (2, 5, and 20 K) deviate from the full scaling relation and are better described solely by the quadratic term [inset of Fig.~\ref{fig:8}(a)], indicating that the intrinsic mechanism alone attributes to the AHE in this regime. Further, a satisfactory fit to the anomalous Hall data of the bare SRO film is obtained using Eq.~\eqref{eq:4} in the temperature range 20-100 K [see Fig.~\ref{fig:8}(a)]. This two-stage behaviour in bare SRO, i.e. intrinsic-only at low temperatures, followed by the onset of skew scattering as well as intrinsic mechanism at higher temperatures signal a temperature-dependent modulation of the Fermi surface \cite{PhysRevB.92.115153}. The fitting of Eq.~\eqref{eq:4} over 20-100 K for bare SRO and 2-100 K for the SRO/SCO bilayer yields $(\alpha, \beta)$ $\approx$ (0.00198, 44.79 S/cm) and (0.0339, 78.4 S/cm), respectively. In case of SRO/SCO bilayer, the increase in $\alpha$ could be due to increased asymmetric scattering arising from the random potentials introduced/present by the interface.

Generally, the anomalous Hall resistivity ($\rho_{xy}$) can be expressed as an integral of the Berry curvature over the occupied Bloch states in momentum space \cite{Fang}, and is therefore sensitive to the Fermi level (E$_F$) position and the spin splitting of the electronic bands. The enhancement of the anomalous Hall coefficient $\beta$ in the SRO/SCO bilayer relative to bare SRO thus reflects a strengthening of the intrinsic Berry curvature contribution. We envisage this to be stemmed from a modification of the Berry-curvature near the Fermi level (E$_F$). The larger magnetization observed in the bilayer suggests an increased exchange splitting of the Ru (4$d$) bands, which could widen the gaps at avoided band crossings in the vicinity of E$_F$ and amplify the net Berry curvature integrated over the Fermi surface \cite{Fang,RevModPhys.82.1539}. Moreover, in the SRO/SCO bilayer, the RuO$_5$ pyramidal coordination at the interface modulates the crystal-field splitting along the c-axis than that in bare SRO, stabilizing the $d_{z^2}$ orbital below $d_{x^2-y^2}$. This interfacial scenario associated with the enhanced intrinsic mechanism ascribing to the AHE bears a striking resemblance to the theoretical report for compressively strained SRO \cite{Tian2021ManipulatingBerryCurvature}, in which crystal-field splitting between the $t_{2g}$ and $e_g$ levels is shown to reorder the Ru $d$-orbital energies with the OOP orbitals ($d_{z^2}$, $d_{zx}$, and $d_{yz}$) lying below the IP ones ($d_{x^2-y^2}$ and $d_{xy}$) and demonstrated to change the sign and the magnitude of the AHC. Therefore, the interface effectively enhances the intrinsic anomalous Hall channel in the SRO/SCO bilayer relative to the bare SRO film.

\section{Conclusion}
In summary, a comparison between the bare SRO film and the SRO/SCO bilayer demonstrates that the SCO overlayer profoundly modifies the magnetic and electronic properties of SRO through interfacial effects. The SRO/SCO bilayer exhibits enhanced saturation magnetization, T$_C$, and PMA, accompanied by suppression of the low-temperature disorder-induced quantum corrections arising from QI and/or EEI below 5 K observed in the bare SRO film, with the electron transport resulting in a Fermi-liquid-like behaviour. Anomalous Hall transport analysis further reveals enhanced intrinsic and skew-scattering contributions in the SRO/SCO bilayer relative to the bare SRO film, indicating substantial modification of the electronic structure attributed to the role of the interface. These experimental findings are discussed envisaging a possible interfacial charge transfer from Cu to Ru facilitated by the RuO$_5$ pyramidal coordination at the interface, in line with Ref.\cite{Li2021HighSpinSCO_SRO} involving a similar SRO/SCO interface. This interpretation is further supported by the enhanced carrier concentration observed in the SRO/SCO bilayer compared to the bare SRO film. Overall, our work elucidates the essential role of the nonisostructural oxide interface in controlling the correlated magnetism and electron transport of SRO thin film.

\begin{acknowledgments}
D.S. and D.P. acknowledge the support from the Institute of Physics, Bhubaneswar.
\end{acknowledgments}

\bibliography{References}

\clearpage
\onecolumngrid

\section*{Supplemental material for Interfacial control of magnetism and electron transport in nonisostructural SRO/SCO heterostructure} 

\author{Digbijaya Palai$^{1,2}$}
\thanks{digbijaya.p@iopb.res.in}
\author{B.~Maharana$^{3}$}
\author{P.~Biswal$^{1,2}$}
\author{Shwetha G.~Bhat$^{4}$}
\author{D.~Nayak$^{2,4}$}
\author{D.~Sahoo$^{6}$}

\author{R.~Soni$^{6}$}
\author{K.~Senapati$^{2,4}$}
\author{Z.~Hossain$^{3}$}

\author{D.~Samal$^{1,2}$}
\thanks{dsamal@iopb.res.in}

\affiliation{$^{1}$Institute of Physics, Bhubaneswar, 751005, India}
\affiliation{$^{2}$Homi Bhabha National Institute, Anushaktinagar, Mumbai, 400094, India}
\affiliation{$^{3}$Department of Physics, Indian Institute of Technology Kanpur, Kanpur, 208016, India}
\affiliation{$^{4}$ Department of Physics, Indian Institute of Science, Bangalore, 560012, India}
\affiliation{$^{5}$Department of Physical Sciences, Indian Institute of Science Education and Research Berhampur, Berhampur, 760010, India}

\subsection*{S1. Temperature dependence of carrier concentration}
Figure S1 show the temperature variation of the carrier concentration obtained from the Hall measurement analysis for the bare SRO film and the SRO/SCO bilayer.

\begin{figure}[!ht]
	\centering
	\includegraphics[width=0.5\linewidth]{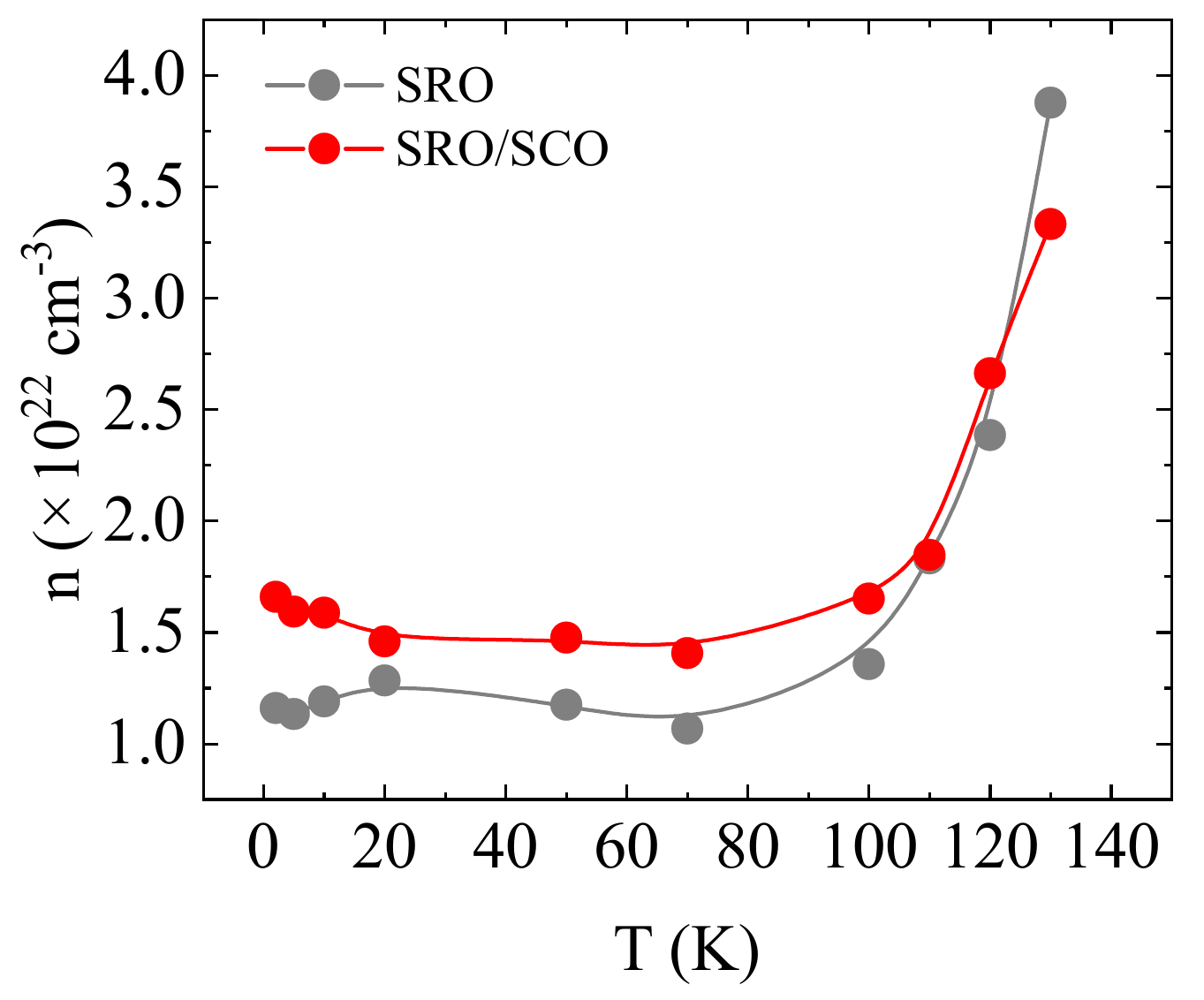}
	\caption{}
	\label{fig:S1}
\end{figure}

\subsection*{S2. Resistivity upturn fitting in the low-temperature regime (2 K $<$ T $<$ 25 K)}
\begin{table}[!ht]
	\centering \setlength{\tabcolsep}{25pt}
	\caption{ The parameters resulting from the low-temperature resistivity fitting using equation (1) for the bare SCO film.}
	\begin{tabular}{ccccc}
		\hline
		SRO & $\rho_{0}$ ($\mu\Omega$ cm) & $\rho_{in}$ ($\mu\Omega$cm/K$^{5}$) & $\rho_{el}$ ($\mu\Omega$cm/K$^{2}$) & $\rho_{l}$ ($\mu\Omega$ cm/K) \\
		\hline
		0 T & 176 & $1.737 \times 10^{-6}$ & 0.0124 & 1.7806 \\
		3 T & 176 & $1.676 \times 10^{-6}$ & 0.0125& 1.7823\\
		\hline
	\end{tabular}
	\label{tab:1}
\end{table}

\begin{table}[!ht]
	\centering \setlength{\tabcolsep}{35pt}
	\caption{ The parameters resulting from the low-temperature resistivity fitting using equation (2) for the SRO/SCO bilayer.}
	\begin{tabular}{ccccc}
		\hline
		SRO/SCO & $\rho_{0}$ ($\mu\Omega$ cm) & $\rho_{in}$ ($\mu\Omega$cm/K$^{5}$) & $\rho_{el}$ ($\mu\Omega$cm/K$^{2}$) \\
		\hline
		0 T & 210 & $7.716  \times 10^{-8}$ & 0.01758 \\
		3 T & 207 & $6.186 \times 10^{-8}$ & 0.01618 \\
		\hline
	\end{tabular}
	\label{tab:table}
\end{table}

\subsection*{S3. Resistivity upturn fitting in the high-temperature regime (25 K $<$ T $<$ 130 K)}
\begin{figure}[!ht]
	\centering
	\includegraphics[width=1\linewidth]{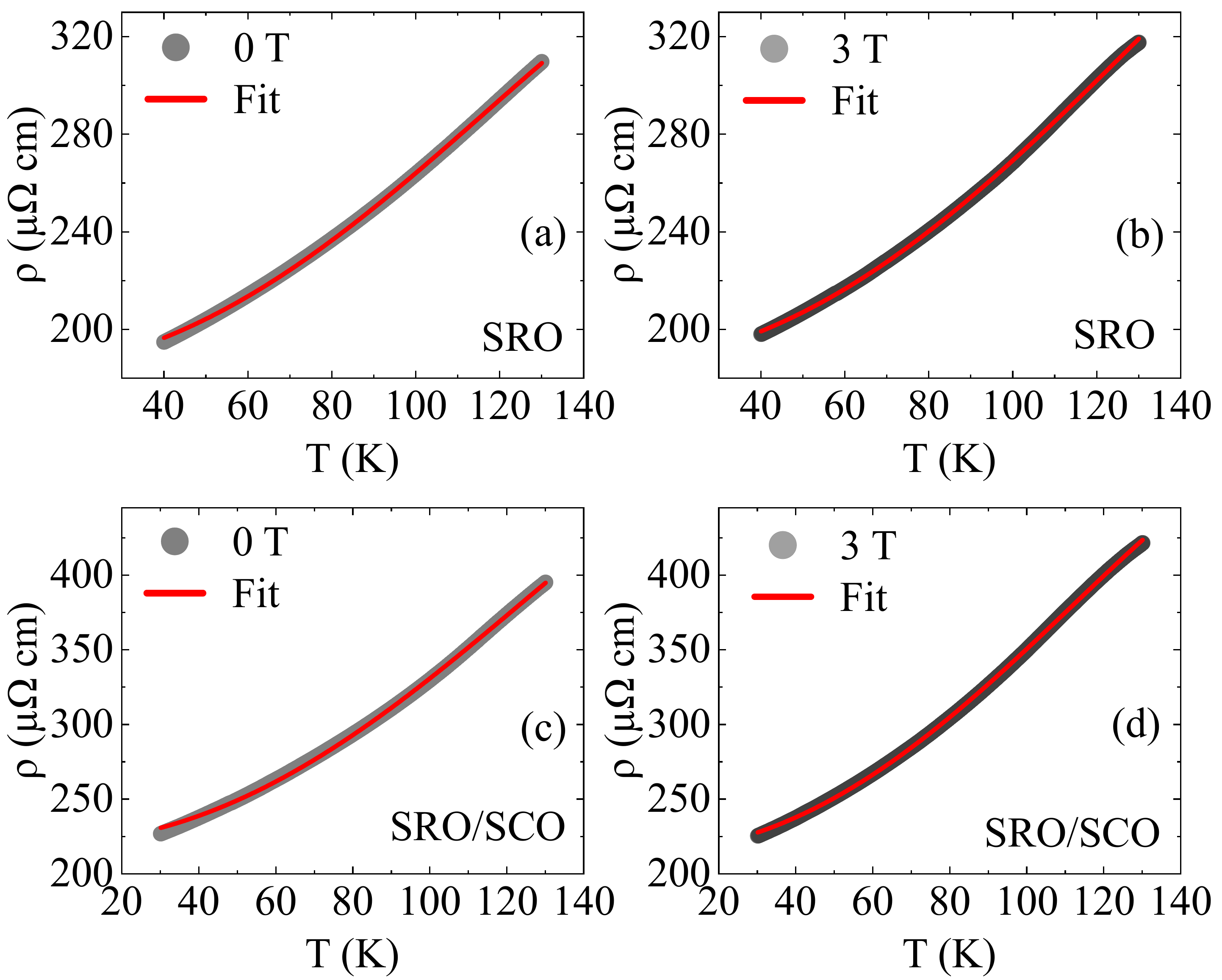}
	\caption{High-temperature (25 K $<$ T $<$ 130 K) resistivity curves fitted with Eq.~\eqref{eq:1} for the bare SRO film [0 T (a), 3 T (b)] and SRO/SCO bilayer [0 T (c), 3 T (d). The scattered symbols correspond to the experimental data and the red solid lines are the fittings to the Eq.~\eqref{eq:1}]}
	\label{fig:S2}
\end{figure}

Beyond the low-temperature quantum correction regime, we analyze the resistivity behaviour in the high-temperature regime (25 K $<$ T $<$ 130 K). Since in this region, the delocalized electron-electron scattering is expected to increase with the increase of temperature, we use the same model given by Eq. ~\eqref{eq:1} to fit the resistivity data of both bare SRO and SRO/SCO bilayer at two different magnetic fields (0 T and 3 T) [see Figs.~\ref{fig:S2}(a)-(d)],

\begin{equation}
	\rho(T) = \rho'_0 + \rho'_{\mathrm{in}}T^5 + \rho'_{\mathrm{el}}T^2
	\label{eq:1}
\end{equation}
where $\rho'_0$ represents the residual resistivity, second term attributes to the presence of electron-phonon and electron-magnon interactions and the last term corresponds to the electron-electron scattering mechanism. As quite evident from the extracted parameters presented in Table-3, we find the contribution from electron-electron scattering coefficients dominates with nearly 10$^6$ time larger than the electron-phonon and electron-magnon interaction coefficients for entire range of temperature with and without the magnetic fields. Note that, $\rho'_{\mathrm{el}}$ is larger in SRO/SCO bilayer than the bare SRO film emphasizes the presence of stronger electron-electron scattering in the SRO/SCO heterostructure, indicating that the electron transport evolves toward a more interaction-driven incoherent scattering regime rather than a disorder-induced localization regime.

\begin{table}[ht]
	\centering \setlength{\tabcolsep}{30pt}
	\caption{ The parameters resulting from the high-temperature resistivity fitting using equation (2) for both bare SRO and SRO/SCO bilayer.}
	\begin{tabular}{ccccc}
		\hline
		Sample & Field  & $\rho_{0}$ ($\mu\Omega$ cm) & $\rho_{in}$ ($\mu\Omega$cm/K$^{5}$) & $\rho_{el}$ ($\mu\Omega$cm/K$^{2}$) \\
		&  &  & $(\times 10^{-10})$ & $(\times 10^{-3})$ \\
		\hline
		SRO & 0 T & 182 & 5.484 & 8.69 \\
		& 3 T & 185 & 4.0386 & 8.81 \\
		SRO/SCO & 0 T & 220 & 8.126 & 11.6 \\
		& 3 T & 214 & 7.425 & 12.05 \\
		\hline
	\end{tabular}
	\label{tab:table}
\end{table}
\clearpage

\subsection*{S4. Anomalous Hall effect in SRO and SRO/SCO samples}
Magnetic field dependence of the anomalous Hall resistance for the bare SRO film [Fig. S3(a) in the grey colored line] and the SRO/SCO bilayer [Fig. S3(b) in the red colored line] measured over the temperature range of 2-130 K, where the sign reversal is observed around 100 K for both samples.

\begin{figure}[!ht]
	\centering
	\includegraphics[width=1\linewidth]{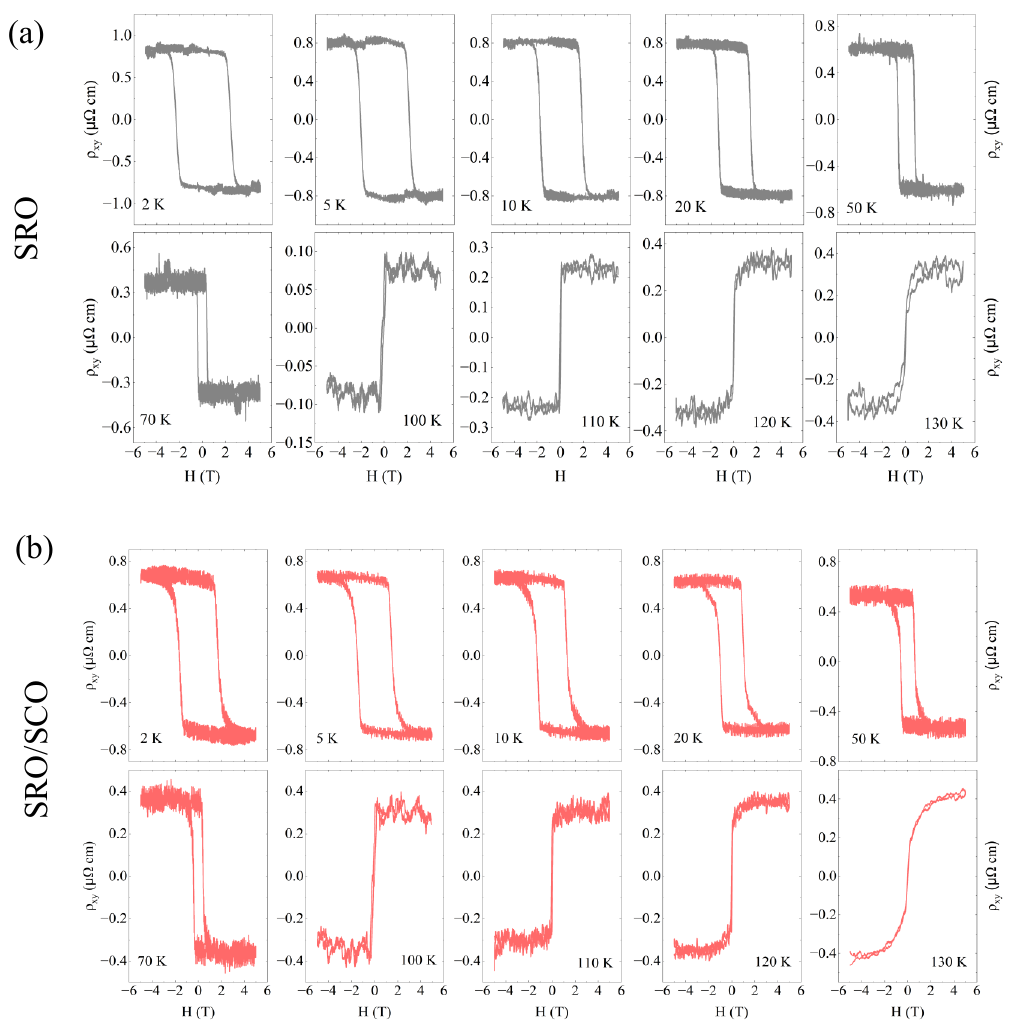}
	\caption{}
	\label{fig:S3}
\end{figure}

\subsection*{S5. Anomalous Hall resistivity versus magnetization}
The anomalous Hall resistivity exhibits a non-monotonous dependance with the magnetization in both the bare SRO film [S4(a)] and the SRO/SCO bilayer [S4(b)], suggesting that the AHE is predominantly governed by the intrinsic Berry-curvature mechanism rather than the side-jump contribution.

\begin{figure}[!ht]
	\centering
	\includegraphics[width=0.8\linewidth]{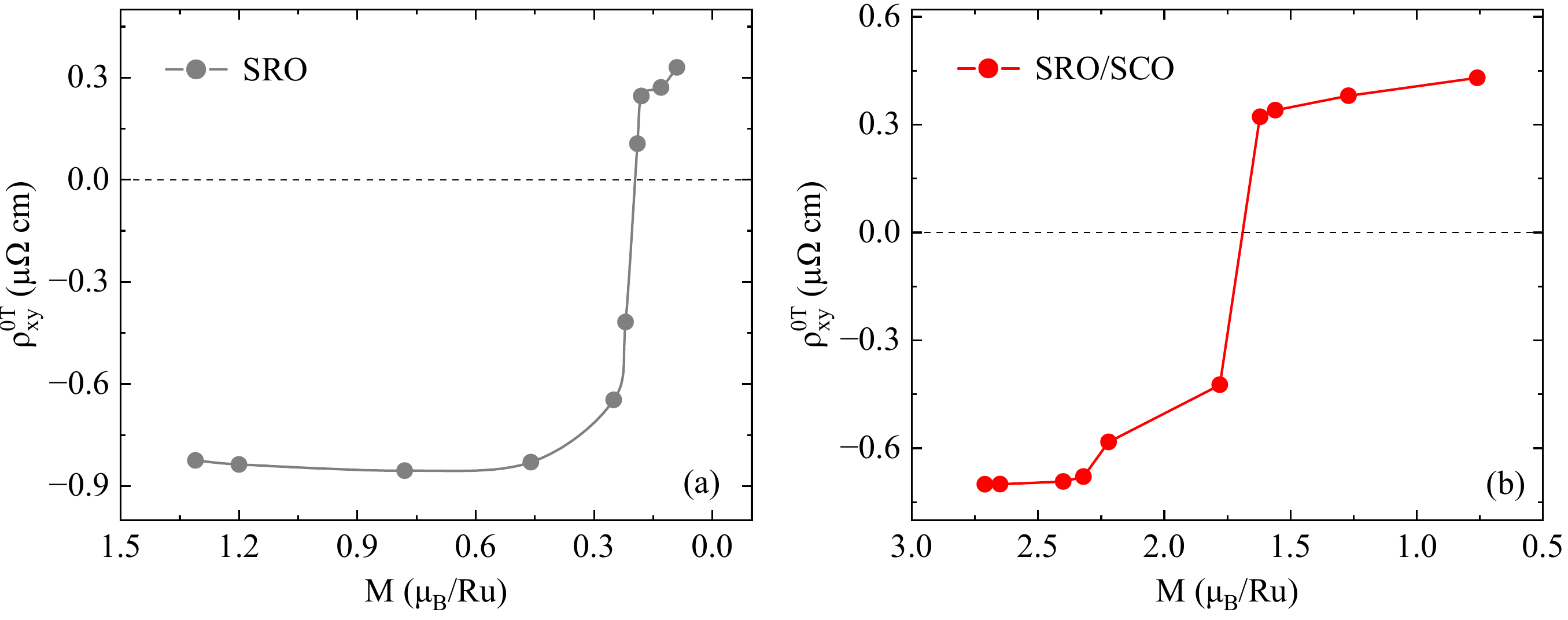}
	\caption{}
	\label{fig:S4}
\end{figure}

\end{document}